\documentstyle[sprocl]{article}
\newcommand{\gtrsim}{\raisebox{-0.3ex}{\mbox{$\stackrel{>}{_\sim}
\,$}}}
\newcommand{\lesssim}{\raisebox{-0.3ex}{\mbox{$\stackrel{<}{_\sim}
\,$}}} 

\begin{document}

\title{SUPERCONDUCTIVITY AND MAGNETISM \\
AT NUCLEAR-MATTER DENSITIES:\\
 AN ASTRONOMICAL CHALLENGE}

\author{M. JAHAN-MIRI}

\address{Institute for Advanced Studies in Basic Sciences \\
 Zanjan 45195, Iran\\E-mail: jahan@iasbs.ac.ir}

\maketitle\abstracts{
We report on a study of the evolution of
magnetic fields of neutron stars, driven by the expulsion of
magnetic flux out of the proton superconducting
core of the star. The rate of expulsion, or equivalently the
velocity of outward motion of
flux-carrying proton-vortices is determined from a solution of
their equation of motion. A determination of the
effective forces on the fluxoids moving through the quantum
liquid interior of neutron stars is however confronted with
many ambiguities about the properties of this special
case of superconductivity in the nature. 
Also, the behaviour of the fluxoids at the core boundary, and
the subsequent evolution of the expelled flux 
within the highly conductive surrounding crust, are other
related issues that have not been so far explored in any
great details.}

\section{Introduction}  
Observational evidence has accumulated that magnetic fields
of neutron stars decay over time scales of tens of million
years or more \cite{TV,PK}. Theoretical modelling of such a field decay calls 
for a study of the expulsion of  
magnetic flux out of the superconducting core of the star, 
which is embedded within the highly conductive crust of
the star. The study would
bear on an understanding of the magnetic and
superconductivity properties of matter under conditions
much different than is usually realised for the terrestrial
superconductors. Some of the novelties might be listed:
\begin{itemize}
\item type I/II proton superconductivity; normal electrons, 
\item ``field cooling'' in an {\em intrinsic} field, originally
        supported by electron currents, 
\item coherent scattering of electrons off fluxoids, 
\item ``transfer'' of flux from proton super-currents to
        electron currents outside, 
\item freezing of fluxoids end points; geometry of the
        expelled flux, and
\item spherical geometry and km size of the superconductor
        and fluxoid lattice.
\end{itemize}
The
velocity of outward motion of flux-carrying proton-vortices
(fluxoids) may be determined from a solution of the Magnus
equation of motion for these vortices \cite{NV,PFC}, taking
into account the various forces that act on them. For the
terrestrial type-II
supercunductors flux movement is usually driven by the
Lorentz force due to an applied transport current.
In contrast, in the interior of a neutron star there are  
other forces acting on the fluxoids. These
include {\em i)} a force due to their pinning interaction
with the moving neutron vortices, {\em ii)} viscous drag force due
to magnetic scattering of electrons, {\em iii)} buoyancy
force, and {\em iv)} curvature force. The derived radial
velocity of the fluxoids at the core-crust boundary would
in turn determine the rate of the flux expulsion out
of the core. The magnetic evolution of the star may be thus
determined by further following the Ohmic dissipation of the expelled
flux within the crust. A detailed description of the modelling 
of such a flux expulsion and field evolution  
and the astrophysical implications of the results of our
computations is reported
elsewhere \cite{JB,J6a,JM9}. Here we give a brief outline
of aspects of the models directly relevant to the 
superconductivity effects. Our aim would be to also highlight
the basic issues, listed above, which require further
theoretical investigations pertaining to the
superconductivity and magnetism in neutron stars.

\section{The Superconductor}
All theoretical models of neutron stars predict a (quantum)
liquid interior (``core'') at densities
($\sim 2.4 \times 10^{14} \ {\rm g~cm}^{-3}$) similar and
above that of the nuclear matter, consisting
of bare neutrons plus an admixture of protons and electrons,
which is surrounded by a solid metallic crust of neutron-rich
nuclei and relativistic degenerate electrons \cite{bpp,R72,PA}.
Neutron stars are ``cold'' objects, having internal
temperatures \(T\sim 10^8 \;({\rm K}) <<
T_{\rm Fermi} \sim 10^{12} \; {\rm K} \). 
For the same theoretical reasons that terrestrial matter is
argued to become superconducting with a transition temperature
\( T_{\rm c}\sim 10^{-3} \; T_{\rm Fermi} \) the neutron stars
are predicted to have superfluid-superconductor interiors
\cite{GK,HGR,TT}. Cooper pairs, in a neutron star, are however
formed due to the long-range attractive part of nucleon-nucleon
interaction among protons, as well as neutrons. The proton
component of the core is usually assumed to form a type-II
superconductor, based on its estimated values of the
penetration depth
$\lambda_{\rm p} \sim 10^{2}$~fm and the coherence length
$\xi_{\rm p} \sim$~few~fm. The possibility of type-I proton
superconductivity, $\xi_{\rm p} > \sqrt{2} \lambda_{\rm p}$,
has been also considered, allowing for the existing 
uncertainities in the values of density, effective mass
of protons, and transition temperature in the core \cite{bpp,Men}. 
However, no attempts have been reported on the modelling of
the field evolution of the star for the case of type-I
superconductivity. The transition to the superconducting state
is believed to occur at a constant field, originally  
present in the core in its normal state \cite{bpp}.  
This should lead to formation of a lattice of proton flux
tubes (fluxoids) which extend some few km across the spherical
core of the star.
Distorted geometries of lines might be also realized due to
the presence of field components other than a pure dipolar
field in the normal state \cite{FR}, which is neglected in our 
models for simplicity.
An interesting feature of this flux trapping
is that, in contrast to the usual requirement for the Meissner
effect, the field need not be expelled in spite of being smaller
than the associated lower critical value $H_{\rm c1} \sim 10^{15}$~G,
which is indeed the case for the typical field strengths of
neutron stars $\sim 10^{12}$~G. The large conductivity of the
matter in the normal state implies dissipation time scales,
for the electron currents in the core, many orders of
magnitudes larger than the time taken for the star to cool
down below its transition temperature $T_{\rm c}$. This is 
argued to imply freezing in of the flux even at field
strengths smaller than the lower critical value \cite{bpp,J6b}.

In the thin layer ($\sim 1$~km) of the crust, surrounding
the superconductor core, matter at the
density range $7\times 10^6 \ ({\rm g~cm}^{-3}) \ 
\lesssim \rho \ \lesssim 2.4 \times 10^{14} \ ({\rm g~cm}^{-3})$
is highly conducting. The estimated Ohmic dissipation time for  
the crust, due to electron scattering by phonons and lattice
impurities is uncertain and lies in the range $10^7$--$10^9$~yr.
Moreover, the unknown geometry of the expelled flux and its 
transport behavior within the crust further complicates the
calculation of its decay time scale \cite{J88,BD}.
The distinctive features of the flux expulsion in the
present case, in contrast to that of the terrestrial
superconductors, 
should be born in mind. Firstly, this is not an expulsion of
the flux associated with an {\em externally} applied field
that might be understood in terms of interaction of the fluxoid
current with the London supercurrent at the surface, and
a readjustment of the latter.
The flux expulsion out of the core of neutron stars requires
electron currents to be induced in the surrounding crust, at
the cost of 
proton supercurrents of the fluxoids being annihilated.  
Furthermore, the fluxoids end points spout out
in a highly conductive surrounding medium. Considering the
spherical geometry of the superconductor, it is not clear
whether the expulsion of the flux is realised only at the
magnetic equator or elsewhere as well. Also, the back
reaction of the flux accumulated in the crust, 
as well as the effect of a surface ``barrier'',   
on the fluxoids motion close to the boundary might have
significant new results. For example, the surface barrier for
the present case, having external fields smaller than
$H_{\rm c1}$, should be an outward declining {\em hill} 
causing the fluxoids to be drained out of the superconductor
\cite{deG}. Such an expulsion mechanism due to surface
effects has not been so far addressed, in the case of neutron
stars; one should see how the corresponding rate of flux
expulsion, if it is realized, compares with that due to the
body forces on fluxoids which are discussed here.

\section{Effective Forces on Fluxoids}
\subsection{Pinning Force}
The rotation of the neutron superfluid implies that 
there exist angular-momentum-carrying neutron-vortex lines
within the core of a neutron star. These are believed to
form a regular lattice parallel to the
rotation axis of the star which is in general tilted with
respect to the magnetic axis, ie. the direction of the
fluxoids. Thus, a neutron vortex might present a
``pinning site'' against a moving fluxoid, vice versa,
should the two structures overlap \cite{MT,Sau,SBM}; some
$10^{31}$ fluxoids and $10^{16}$ vortices should be,
typically, present in a neutron star.
In the steady-state, the n-vortices must be co-rotating
with the charged component of the star, including the
lattice of the p-fluxoids, at a rate $\Omega$. 
For a given rate $\dot \Omega$ of spinning
down of the star the radial velocity $v_{\rm n}$ of
the n-vortex outward motion, at the core boundary, would be 
\begin{eqnarray}
v_{\rm n} &=& -{ R_{\rm c} \over 2} { {\dot \Omega } \over
\Omega_{\rm s} } \approx -{ R_{\rm c} \over 2} { {\dot \Omega}
\over \Omega }
\end{eqnarray} 
where $R_{\rm c} (\sim 9 \times 10^5$~cm) is the radius of
the stellar core, and $\Omega_{\rm s} (\sim\Omega)$
is the superfluid rotation rate.
A superfluid normally spins down while maintaining a positive 
rotational lag $\omega (\equiv \Omega_{\rm s} - \Omega >0$)
with its vortices. Hence an outward radial Magnus force
$F_{\rm M} = \rho_{\rm s} \kappa R_{\rm c} \omega$ would 
act on the vortices, per unit length at the core boundary,
where $\rho_{\rm s}$ is the neutron-superfluid density, and
$\kappa= 2 \times 10^{-3} {\rm cm}^2 {\rm s}^{-1}$ is the
vorticity of a vortex line \cite{Sau}.

The Magnus force on n-vortices has been usually assumed to be
balanced by the viscous forces being primarily caused by
the scattering of electrons off the magnetized cores of
these vortices in the interior of a neutron star \cite{Sau}.
However in presence of the pinning forces it turns out that
the latter would be the dominant force which will effectively 
equate with the Magnus force, resulting in a net zero force on
the n-vortices. This force balance condition may be used to
calculate the
magnitude of the pinning force on n-vortices, as well as its
direction which could be, in principle, in either (inward
or outward) radial directions depending on
the relative motion of the vortices and fluxoids.

The effective pinning force per unit length of each fluxoid
may be, in turn, evaluated by noting the equality of the mutual
pinning forces between fluxoids and vortices at each pinning
site, and then balancing the total pinning force communicated
between the two lattices of vortices and fluxoids. However, 
at any given {\it instant} of time only a small fraction
of the vortices would be directly interacting with the
fluxoids. The remaining much greater fraction of them (of
the order of the ratio of an inter-fluxoid spacing to
the size of a pinning interaction region; see below) should
reside in the inter-fluxoid spacings.
Thus the total Magnus force acting {\em continuously on all}
the vortices may, or may not, be communicated instantaneously
to {\em all} the fluxoids, depending on the assumed rigidity
of the n-vortices and other considerations. We omit a detailed
discussion of
the derivation of the pinning force on fluxoids for the 
different possibilities and write down the final results. 
The pinning force $F_{\rm n}$ acting on a fluxoid, per unit
length, is derived as \cite{JM9}, either 
\begin{eqnarray}
F_{\rm n} &=& { n_{\rm v} \over n_{\rm f}} F_{\rm M} \approx 
	2 \phi_{0} \ \rho_{\rm s} \ R_{\rm c} \ 
        {\Omega(t) \ \omega(t) \over B_{\rm c}(t)} 
        = 5.03  { \omega_{-6} \over P_{\rm s} B_{8}} \ \  
	{\rm dyn cm}^{-1}
\end{eqnarray}
where $n_{\rm v} = { 2 \Omega_{\rm s} \over \kappa}$ and $n_{\rm f}=
{B_{\rm c} \over \phi_{0}}$ are the number densities per unit cross
section area of the vortices and the fluxoids, respectively,
$\phi_{0}=2 \times 10^{-7} {\rm G\ cm}^2$ is the
magnetic flux carried by a fluxoid, $B_{\rm c} = 10^8 B_{8}$ is the
strength of the core field in units of G, $\omega_{-6}$ is the
superfluid lag $\omega$ in units of $10^{-6} {\rm rad \ s}^{-1}$, 
and $P_{\rm s}$ is the spin period in units of s.
Notice that the sign of $\omega$ determines the sign of $F_{\rm n}$
for which, as well as for the other forces discussed below, the
outward direction will be reckoned as the positive sense.
Or, else,
\begin{eqnarray}
F_{\rm n} &=& { d_{\rm P} \over d_{\rm f}} \left( { n_{\rm v} \over 
        n_{\rm f}} F_{\rm M} \right)  
        =2.59 \times 10^{-4}  { \omega_{-6} \over 
          P_{\rm s} B_{8}^{1/2}} \ \ {\rm dyn \ cm}^{-1}
\end{eqnarray}
where $d_{\rm f} = 2.3 \times 10^{-7} B_{8}^{-{1\over2}} $~cm is
the inter-fluxoid spacing, and $d_{\rm P}$ is the effective size of a
pinning interaction region around each fluxoid. A value of $d_{\rm P} =
\lambda_{\rm p}^* = 118$~fm has been used for the assumed
magnetic pinning mechanism (see below), where $\lambda_{\rm p}^* $
is the effective London length of the proton superconductor, being 
also a length scale for the spread of the magnetic field
of a neutron vortex line.

\vspace{.5 cm}
\noindent
{\bf Critical lag:} \\ 
The magnitude of the force which could be exerted {\em at each
intersection} by a vortex on a fluxoid, and vice versa,
is limited by a maximum value $f_{\rm P}$ corresponding
to the given strength of the pinning energy $E_{\rm P}$ and the
finite length scale of the interaction $d_{\rm P}$; namely
$E_{\rm P} = f_{\rm P} d_{\rm P}$.  This implies a maximum
limiting value that the Magnus force on the n-vortices
could achieve, 
corresponding to a maximum critical lag $\omega_{\rm cr}$ 
which is given as
\begin{eqnarray}
	\omega_{\rm cr} &=& 1.59 \times 10^{-6} \ B_{8}^{1/2} 
	\ \ {\rm rad \ s}^{-1}
\end{eqnarray}
The critical lag
is the magnitude of the lag when radial velocities
of fluxoids and vortices are different; ie. 
\hbox{ $ \omega = \omega_{\rm cr} $} or \hbox{ $ \omega = -
\omega_{\rm cr} $} when the vortices move faster or slower than
the fluxoids, respectively. However, during a co-moving phase
when the force communicated between a vortex and a
fluxoid at each pinning point is less than its maximum value,
$f_{\rm P}$, the lag might have any value within the range
\hbox{ $ - \omega_{\rm cr} < \omega < \omega_{\rm cr}$}.
The pinning energy arising from the magnetic interaction of a
fluxoid-vortex pair is estimated to be
$E_{\rm P} \sim 10^{-5}$~ergs. A different estimate for the
pinning energy, due to the proton density perturbation, gives
a smaller value of $E_{\rm P} \sim 5 \times 10^{-7}$~ergs
\cite{Sau,J91}. However, both mechanisms result in similar
values for the pinning force, hence similar $\omega_{\rm cr}$,
since the interaction length, $d_{\rm P}$, for the latter
($=\xi_{\rm p}$) is smaller than for the magnetic
interaction ($=\lambda_{\rm p}^*$) by about the same ratio
as the inverse of the pinning energies.

\subsection{Drag force}
An isolated fluxoid moving through the normal degenerate electron
gas in the core of a neutron star is subject to the viscous drag force
of the electrons scattering off its magnetic field. The viscous drag
force, per unit length of a fluxoid, is estimated to be
\cite{HRS,J87}
\begin{eqnarray}
	\vec{F_v} &=& - {3 \pi \over 64} { n_{\rm e} e^2 \phi_0^2 
        \over E_{\rm F_{\rm e}} \lambda_{\rm p}} 
        { \vec{v_{\rm p}} \over c}   
        = -  7.30 \times 10^7 \ \vec{v_{\rm p}} \ \ 
	{\rm dyn \ cm}^{-1} 
\end{eqnarray}
where $v_{\rm p}$ is the velocity of the outward radial motion of
the fluxoids in units of ${\rm cm \ s}^{-1}$,  
$n_{\rm e} = 3.  \times 10^{36} {\rm cm}^{-3}$ is the number
density of the electrons, and $E_{\rm F_{\rm e}}= 88 $~MeV is the
electron Fermi energy, corresponding to a total density
$\rho = 2 \times 10^{14} {\rm g\ cm}^{-3}$, and a neutron number 
density $n_{\rm n} = 1.7 \times 10^{38} {\rm cm}^{-3}$ in the core.

The expression for $F_v$ in Eq.~5 is derived based on the
assumption of independent motions for single fluxoids. However,
for the typical conditions in the interior of a neutron star the
lattice of fluxoids might be, as a whole (or at least as bundles
consisting of not less than ten million fluxoids), ``frozen-in''
the electron gas \cite{HRS}. This is
because the mean distance between the successive magnetic
scattering of electrons by fluxoids turns out to be many
orders of magnitudes smaller than the mean free path of the
electrons due to other events, and also that the deflection
angle at each scattering event is very small. A detailed
treatment of the coherent electron scattering by the fluxoid
lattice has been shown to indeed require an almost
{\em zero relative velocity} between the electrons and the
lattice. Therefore, the flux expulsion out of the core might
be {\em prohibited} altogether except if electron-current
loops across the core-crust boundary is realized.
Uncertainties about the true
distribution of the magnetic flux and the correct value of the
conductivity in the crust, and also the possibility of a 
mechanical failure of the solid crust due to a build-up of
the magnetic stresses, however, obscure any definite conclusion 
to be drawn \cite{J87}.

Moreover, there are other reasons to suspect the above
suggested frozen-in approximation for the fluxoids as a whole   
\cite{DCC,R95}. For example,
the finite volume of the fluxoid lattice and also the influence
of the superconductor boundary effects on the motion of the
fluxoids which have not been included in the above mentioned
studies of
the coherent scattering could as well have significant new 
consequences. In addition, the motion of the incompressible
electron fluid in the interior of a neutron star has been argued
to be divergence free \cite{GR}. Any motion of the fluxoids
along with the electrons, in the frozen-in approximation, must
be, therefore, of the same (divergence-free) nature. This is
however impossible for the uniform lattice of fluxoids during
its outward motion since the lattice constant keeps changing.
Hence, a compromise between the
flux freezing and the divergence-free motion of the electrons has
to be worked out if any flux expulsion is to be accounted for.

Given the above uncertainties, as well as the lack of any other
definite prescription for calculating the drag force due to the
electron scattering, we use a value of $F_v$ as given by Eq.~5 
in our models. This choice is also supported by noticing that
a tentative expulsion time
scale derived for the case of coherent scattering, as implied by
the Hall drift of the flux at the base of the crust, 
turns out to be similar to that based on the single fluxoid
approximation \cite{J88,J91}.

\subsection{Buoyancy force} 
The buoyancy force on fluxoids in a neutron star arises for reasons
analogous to the case of macroscopic flux tubes in ordinary stars.
Because flux tubes are in pressure equilibrium with their 
surrounding the excess magnetic pressure causes a
deficit in the thermal pressure, and hence in the density, of the
plasma inside a flux tube which make the tube to become buoyant.
The radially outward buoyancy force $F_{\rm b}$ on a fluxoid,
per unit length, can be expressed as \cite{MT,J87} 
\begin{eqnarray}
	F_{\rm b} &=& \left( { \phi_{0} \over {4 \pi 
	\lambda_{\rm p}} }
        \right)^2 { \ln (\lambda_{\rm p} / \xi_{\rm p}) \over R_{\rm
        c}  }  
                  = 0.51 \ \ {\rm dyn \ cm}^{-1}
\end{eqnarray}

\subsection{Curvature force} 
The tension of a vortex line (such as a fluxoid) implies that
a curved geometry of the line would result in a restoring force,
the curvature force, which tries to bring the line back to
its minimum energy straight configuration. The concavely directed
curvature force $F_{\rm c}$, per unit length, on a vortex having a
tension $T$ and a curvature radius $S$ is given as \(
F_{\rm c} = { T/ S } \). An outward moving fluxoid might be
bent outward and becomes subject to an inward curvature force,
since its end points are frozen in at the highly conductive
bottom of the crust. Thus, for a fluxoid with a tension
$T_{\rm p} =\left( { \phi_{0} / {4 \pi \lambda_{\rm p}}
}\right)^2 { \ln (\lambda_{\rm p} / \xi_{\rm p})}$ 
the curvature force, per unit length, would be given as
\cite{HRS,DCC}
\begin{eqnarray}
F_{\rm c} &=& - {R_{\rm c} \over S} 
\left( { \phi_{0} \over {4 \pi \lambda_{\rm p}} }\right)^2 
{ \ln (\lambda_{\rm p} / \xi_{\rm p}) \over R_{\rm c}}  
         \equiv - {R_{\rm c} \over S} F_{\rm b}
         = - 0.35  \ \ {\rm dyn \ cm}^{-1},
\end{eqnarray}  
using a value of
${R_{\rm c} \over S} \sim \ln{2}$ for an assumed spatially
uniform distribution of the fluxoids.

The electron currents at the bottom of the crust are however subject to
diffusion processes, the rate of which would set a maximum
limiting speed $v_{\rm max}$ for the fluxoids end points
to move. We, therefore, assume that whenever
\( v_{\rm p} < v_{\rm max} \) the fluxoids remain straight and
{\em no} curvature force will be acting on them ($F_{\rm c}=0$),
since their end points are also able to move with the same speed.
In the opposite case, when \( v_{\rm p} \geq v_{\rm max} \), 
the fluxoids would be bent outward and the force
$F_{\rm c}$ as in Eq.~7 would be effective.
This maximum drift velocity $v_{\rm max}$ for the magnetic
flux in the crust is estimated, based on the Ohmic diffusion
alone, as
\begin{eqnarray}
        v_{\rm max} &\sim& {R \over \tau }   
        = 3.18 \times 10^{-9}  \left( { \tau \over
        10^7 {\rm yr} } \right)^{-1}  \ \ {\rm cm \ s}^{-1}
\end{eqnarray}
where $R=10^6$~cm  is the radius of a neutron star, and 
$\tau $ is the assumed time scale, in units of yr,
for the decay of the magnetic field in the crust. A larger value
for $v_{\rm max}$ may be expected if the Hall drift of the magnetic
flux at the bottom of the crust is also taken into account.

On the other hand, the collective rigidity of the fluxoids
lattice due to mutual repulsive forces between them might 
react to a deformation \cite{J91}. The force $F_{\rm c}$
associated
with even a piece of the lattice of a size of an inter-vortex
spacing (including some $10^7$ flux lines)
would be, accordingly, so large that any bending of the
lattice is effectively prohibited.  The velocity of the fluxoids
would then be constrained at all times by the condition \(
v_{\rm p} \leq v_{\rm max} \). We use alternative models to
test this possibility as well.

\section{The Models} 
The steady-state radial motion of a fluxoid, in the region of
interest, is thus determined from the balance equation for all
the radial forces acting on it, per unit length, that is
\cite{NV,PFC}:
\begin{eqnarray}
	F_{\rm n} +F_v +F_{\rm b} +F_{\rm c} &=& 0
\end{eqnarray}
which may be rewritten in the form
\begin{eqnarray}
 \alpha \ {\omega_{-6} \over P_{\rm s} B_{8}} - \beta \ v_{\rm p_7} + 
\delta = 0 
\end{eqnarray}
where parameters $\alpha$, $\beta$, and $\delta (\equiv F_{\rm b} +
F_{\rm c})$ would have different values for the alternative 
models that we have considered, and $v_{\rm p_7}$ is the
fluxoid velocity $v_{\rm p}$ in units of
$10^{-7}$~cm~s$^{-1}$.  Recall that $\omega_{-6}$, which is the value
of $\omega$  in units of $10^{-6} \ {\rm rad \ s}^{-1}$, might have
either positive or negative values, as is also the case with $\delta$
in some of the models.

This single equation includes two unknown variables $\omega$ and
$v_{\rm p}$,
and represents the {\em azimuthal} component of the Magnus
equation of motion for the proton vortices.
No radial Magnus force acts on the fluxoids, hence the right hand
side is set equal to zero, because of the assumed co-rotation of
the fluxoids with the proton superconductor.  There exist however
additional restrictions on the motion of the fluxoids which help  
to fix the value of one of the variables and solve Eq.~10 
for the other.  Namely, for a co-moving state \( v_{\rm p} = v_{\rm
n}\) is given and $\omega$ could be determined. In contrast,
when $v_{\rm p}(\neq v_{\rm n})$ is unknown $\omega$ is given as
\( \omega =\omega_{\rm cr} \) or \( \omega = - \omega_{\rm cr} \)
for \( v_{\rm p} < v_{\rm n} \) or \( \ v_{\rm p} > v_{\rm n} \),
respectively. Furthermore, inspection of Eq.~10 indicates that it
admits {\em one and only one} of the three different solutions, 
for the given values of $v_{\rm n}, B_{\rm c}, $ and $P_{\rm s}$
at any time, namely 

\begin{tabular}{llc}
$\omega$ = $\omega(v_{\rm p}=v_{\rm n})$ & \hspace{1.2 cm} iff 
 \hspace{1.2 cm}& $- \omega_{\rm cr} < \omega <\omega_{\rm cr}$      \\
$v_{\rm p}$ = $v_{\rm p}(\omega=\omega_{\rm cr})$ & \hspace{1.2 cm}
iff  \hspace{1.2 cm} & $v_{\rm p} < v_{\rm n}$                       \\
$v_{\rm p}$ = $v_{\rm p}(\omega=-\omega_{\rm cr})$ &  \hspace{1.2 cm}
iff  \hspace{1.2 cm} & $v_{\rm p} > v_{\rm n}$                         \\
\end{tabular}
 
\noindent
The rate of the
flux expulsion out of the core, $\dot B_{\rm c}= -{2 \over R_{\rm c}}
B_{\rm c} v_{\rm p}$, and the evolution of the stellar surface field
$B_{\rm s}$
(with a decay rate $\dot B_{\rm s} = -{{ B_{\rm s} - B_{\rm c} }
\over \tau}$) are hence uniquely determined from the above
force balance equation, given the spin evolution of the star which
determines (Eq.~1) the vortex velocity $v_{\rm n}$ at any time.

\subsection{The results}
The computed time evolution of $v_{\rm p}$ and $\omega$ are shown in
Fig.~1, together with $v_{\rm n}$ and $\omega_{\rm cr}$, as is
predicted in one of the models. Characteristically similar results as
in Fig.~1 are obtained for the other tested models as well.
The fluxoids motion in Fig.~1 is seen to follow three
evolutionary
\begin{figure}[h]
\vbox to4.5cm{\rule{0pt}{4.5cm}}
\includegraphics{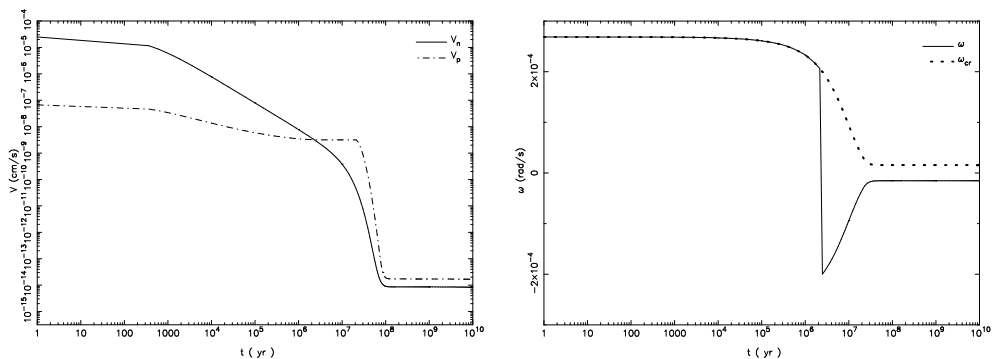}
\caption
{The {\em right} panel shows the predicted time
	evolution of the lag $\omega$ and its critical value
        $\omega_{\rm cr}$ in a solitary neutron star according to 
        one of our simulated models. The {\em left} panel
        shows the corresponding
        evolution of the velocities of the fluxoids ${\rm V}_{\rm p}$,
        and the vortices ${\rm V}_{\rm n}$.  Initial values of $B_{\rm
	s}=10^{12.5}$~G, $B_{\rm c} = 0.9 B_{\rm s}$, and a value of
        $\tau = 10^7$~yr have been used.}
\end{figure}
phases in which they move slower, together, and faster than the
vortices, successively. Transitions between these
successive evolutionary phases occur because of the reduction in
$v_{\rm n}$ ($\propto \dot \Omega_{\rm s}$) as well as the increase
in $P_{\rm s}$; a final co-moving phase of fluxoids and
vortices might also occur for
some choices of the initial conditions.

The predicted evolution of the core and surface fields
according to the same model is given in Fig.~2.
A substantial decrease in the core field occurs at a time $ t \gtrsim
10^7$~yr, which is expected for the typical average values of $v_{\rm
p} \lesssim 10^{-8}\ {\rm cm \ s}^{-1}$ during the earlier times,
because \( \frac{\dot B_{\rm c}}{B_{\rm c}} =
\frac{v_{\rm p}}{R_{\rm c}} \) implies that a time period \( \Delta
t \sim \frac{R_{\rm c}}{v_{\rm p}} \) is needed for a major reduction
in the core field to
\begin{figure}[h]
\vbox to4.5cm{\rule{0pt}{4.5cm}}
\includegraphics{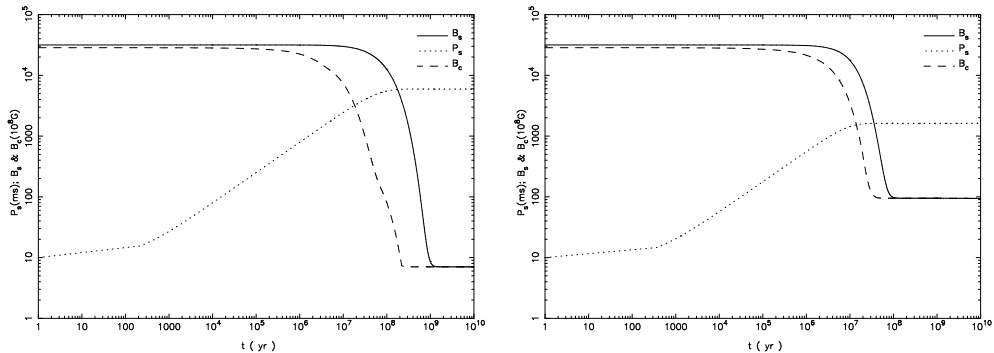}
\caption
{The predicted time evolution of the strength of the
	magnetic field in the core $B_{\rm c}$ and at the surface
	$B_{\rm s}$, and the spin period $P_{\rm s}$ in a solitary
        neutron star according to the same model as in Fig.~1.
        The {\em right} panel
        {\bf (a)} is for an assumed value of $\tau = 10^7$~yr and
        corresponds to the results in Fig.~1, while the {\em left}
        panel {\bf (b)} is for $\tau = 10^8$~yr.}
\end{figure}
occur.
However, because of the very small
magnitude of $v_{\rm p}$ (although $\gtrsim v_{\rm n}$) and also
the reduced value of $B_{\rm c}$ at later times $B_{\rm c}$ does
not change, substantially, afterwards.
The surface field $B_{\rm s}$ responds
to the change in $B_{\rm c}$ on
the assumed decay time scale $\tau$ of the crust. The nontrivial role 
of the stellar crust in these field evolution models may be seen by
comparing Fig.~2a with Fig.~2b, where values of $\tau = 10^7$,
and $10^8$~yr have been used, respectively. A larger value of
$\tau$ tends to maintain the initial $B_{\rm s}$, hence a larger 
$\dot P_{\rm s}$ as well as a larger $v_{\rm n}$, over a more extended
period of time.  Consequently, smaller final values of $B_{\rm c}$
and $B_{\rm s}$ are predicted for the larger assumed values of
$\tau$, as is seen in Fig.~2. 

Further consideration of the predicted time evolution of the
forces on fluxoids, contrasted with the field evolution, 
reveals that the dominant
``driving'' force for the flux expulsion is the buoyancy force
which is positive throughout the evolution.
Accordingly, the overall role of the pinning force in 
the field decay of neutron stars turns out to be more like a
``brake'', preventing the flux from being otherwise
{\em expelled too rapidly}.
Our conclusion about the braking role of the pinning force is,
however, new and in contradiction with the earlier claims 
\cite{SBM,DCC}.

\end{document}